\documentclass[]{article}
\usepackage{amsmath,epsfig,spconf}%

\usepackage{xcolor}
\usepackage{graphicx}
\usepackage{url}
\usepackage{booktabs}
\usepackage{multirow}
\usepackage{bbm}
\usepackage{cite}
\usepackage{stmaryrd}
\usepackage{float}
\usepackage{amsfonts}
\usepackage{amssymb}
\usepackage{amsbsy}
\usepackage{amsmath}
\usepackage{graphicx} 
\usepackage{subcaption}
\usepackage{enumitem}

\usepackage{hyperref}
\hypersetup{
    bookmarks=true,         
    unicode=false,          
    pdftoolbar=true,        
    pdfmenubar=true,        
    pdffitwindow=false,     
    pdfstartview={FitH},    
    pdftitle={},    
    pdfauthor={},     
    pdfsubject={},   
    pdfcreator={},   
    pdfproducer={}, 
    pdfkeywords={}{causality}, 
    pdfnewwindow=true,      
    colorlinks=true,       
    linkcolor={blue},          
    citecolor=blue,        
    filecolor=blue,      
    urlcolor=blue           
}

\begin{document}


\title{A deep network approach to multitemporal cloud detection} 

\name{Devis Tuia$^1$, Benjamin Kellenberger$^1$, Adrian P\'erez-Suay$^2$, Gustau Camps-Valls$^2$
\thanks{DT and BK acknowledge the Swiss National Science Foundation (no. PP00P2-150593). APS and GCV acknowledge the support by the Spanish Ministry of Economy and Competitiveness (MINECO-ERDF, TIN2015-64210-R and TEC2016-77741-R projects), the EUMETSAT Contract No. EUM/RSP/SOW/14/762293, and by the European Research Council (ERC) under the ERC-CoG-2014 SEDAL under grant agreement 647423.}\vspace*{-3mm} 
}

\address{$^1$ Laboratory of GeoInformation Science and Remote Sensing, Wageningen University, The Netherlands\\ 
$^2$Image Processing Laboratory, Universitat de Val\`encia, iSpain\\
\small{\texttt{\{devis.tuia,benjamin.kellenberger\}@wur.nl}, \texttt{\{gustau.camps,adrian.perez\}@uv.es}} 
\vspace*{-3mm} }

%
\maketitle
\begin{abstract}
We present a deep learning model with temporal memory to detect clouds in image time series acquired by the Seviri imager mounted on the Meteosat Second Generation (MSG) satellite. 
The model provides pixel-level cloud maps with related confidence and propagates information in time via a recurrent neural network structure. With a single model, we are able to outline clouds along all year and during day and night with high accuracy. 
\end{abstract}


\section{Introduction}

\begin{figure}[!t]
\centerline{\includegraphics[width = .75\linewidth]{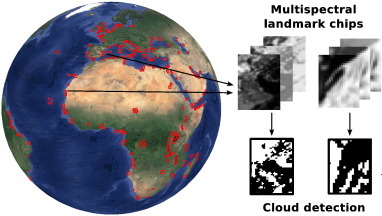}}
\caption{Principle of cloud detection over landmark locations.}\label{fig:globe}
\end{figure}

Cloud detection is paramount for a wide amount of tasks exploiting remote sensing optical data. For example, {in the case of ground-based targets,} clouds mask the Earth's surface and provide incorrect reflectance values. In this case, they must be excluded (as other missing data) and replaced before the image can be further used: common strategies include temporal interpolation along a time series~\cite{Pel16} or spatial geostatistics~\cite{Zha09}. Another example is weather forecasting, where geostationary satellite missions {like} the Meteosat Second Generation (MSG) acquire wide swath images {repetitively} covering the same portions of the globe, making fine registration of the single acquisitions compulsory for image navigation and geometric calibration~\cite{Just00}: such registration is generally performed by using landmark points defined all over the globe (Fig.~\ref{fig:globe}). In this case, it is compulsory to know when such landmarks are covered by clouds, since the registration may become inaccurate~\cite{PEREZSUAY201754}. In this paper we tackle this second scenario.

We consider the problem of cloud detection over a time series covering a landmark repetitively imaged by the Seviri instrument. In~\cite{PEREZSUAY201754}, the task is approached at the single pixel level and by a classical `feature extraction, then classification' approach, where a number of spectral and contextual features are first extracted at the image level and then each pixel is used in a committee of experts formed by dedicated Support Vector Machine classifiers for single pixel cloud classification. The system runs four separate models, specializing to four different periods of the day characterized defined by specific sun zenith angles.

In this paper, we approach the problem from a similar perspective, since we also design a system first extracting contextual features from the spectra and then deploying a pixel-classifier. However, we propose a \emph{single}, \emph{end-to-end} framework, where features and classifier are learned by a deep Convolutional {N}eural {N}etwork {(CNN)}. {CNNs} have become a prominent framework of interest in remote sensing~\cite{Zhu17}: their characteristic to learn naturally contextual features at multiple scales~\cite{Vol17} makes them a promising framework for pixel-level, dense semantic segmentation (in our case) of infrared images. We limited our inputs to the infrared bands of the Seviri instrument (we omitted the RGB bands) to have an input space that is informative, both during day and night, and thus design a single CNN classifier providing the cloud mask for the image, independently from the specific time of acquisition.

Despite the good results provided by such approach{es}, proceeding this way ignores the temporal structure of the data. Note that the Seviri images are acquired every 15 minutes so temporal correlation about the clouds presence should be also informative to improve performance. We therefore expand the model to include temporal dependencies, by unwrapping a sequence of identical CNNs (with  shared weights) and allowing the prediction obtained at the previous time step to influence the next one. By doing so, we inject temporal memory in the CNN, {which as such becomes} a Recurrent Neural Network (RNN)~\cite{888}.

\section{Unwrapped {CNN} with temporal memory}

This section introduces the CNN model and the strategy followed to inject temporal memory in the model. 

\subsection{Core model: the modified hypercolumn CNN}\label{sec:hc}

\begin{figure}[!b]
\includegraphics[width = .95\linewidth]{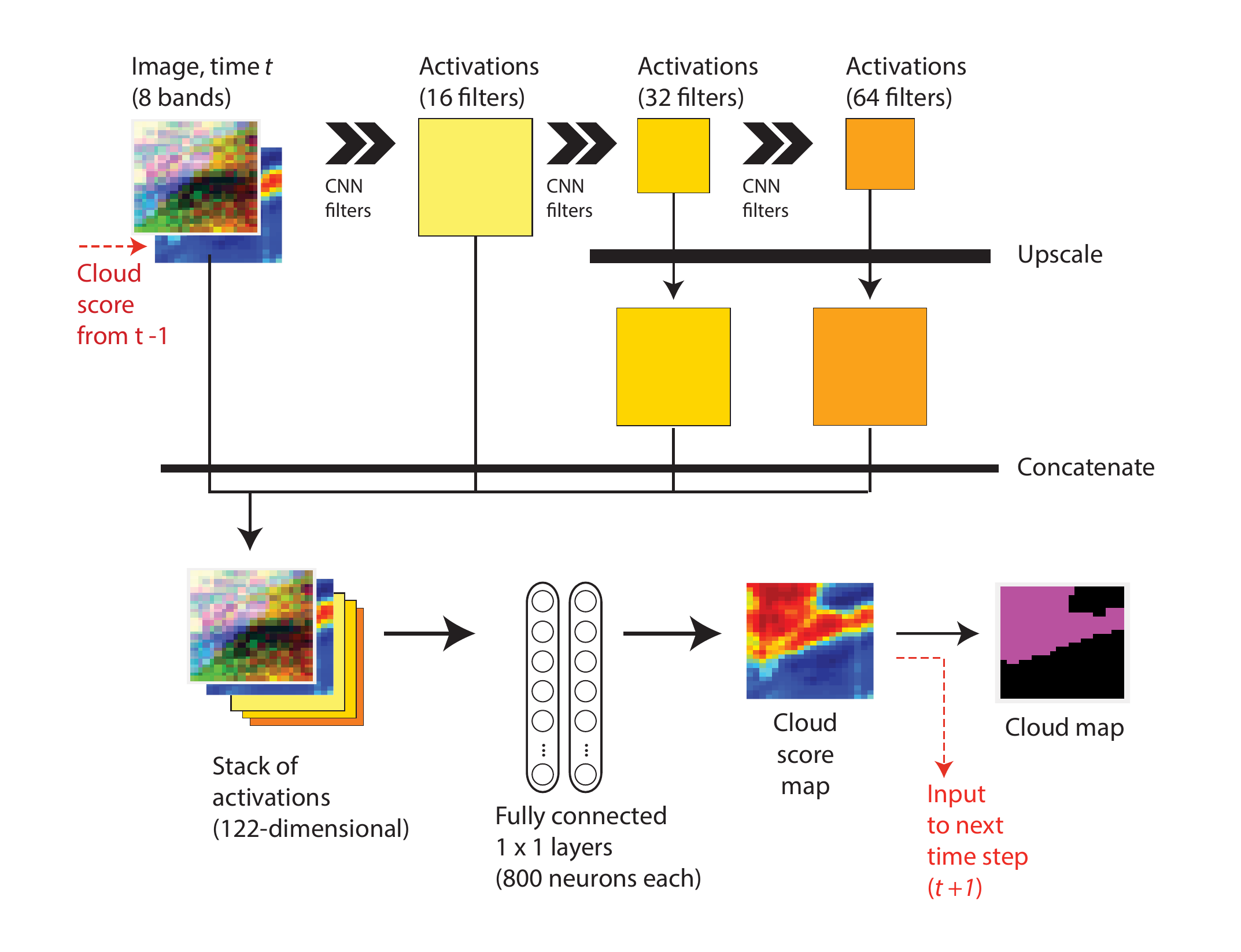}
\caption{Hypercolumn CNN used as the core architecture. Red arrows depict the temporal connections added to the base architecture.
}\label{fig:archi}
\end{figure}

Figure~\ref{fig:archi} presents the core network architecture used in this study. It follows the hypercolumn architecture originally presented in \cite{hariharan2015cvpr} with the modifications proposed in \cite{maggiori2017tgrs,marcos2018jisprs}.

The main trunk of the architecture (top part of Fig.~\ref{fig:archi}) consists {of} a standard classification network. Such {a} network learns a set of filters organized in three blocks, each one separated by nonlinearities (using the ReLU unit), batch normalization, and spatial downsampling (max-pooling) operations (see~\cite{Vol17} for details about these operations).

Differently from a classical classification network, the following fully-connected layer is removed and the activations are used instead: each activation set (corresponding to the result of convolving the input with a set of learned filters) is upsampled at the original image resolution and concatenated to it. As a result of this upsampling-plus-concatenation phase, we retrieve a 3D tensor of the same size as the original image in the two first dimensions ($r \times c$), and with as many features as the sum of the original bands plus the number of filters learned by the CNN in the third dimension. Proceeding this way mimics a contextual feature extraction phase, since it results in a set of hierarchical features for each pixel.

This tensor is then used in a fully convolutional way~\cite{long2015cvpr}, i.e. to perform dense segmentation providing the cloud score at every image location. To do so, we use $1 \times 1$ convolutions mapping into binary class scores. This is equivalent to learning a multilayer perceptron (MLP) model. Since the proposed model involves only standard convolutions operations, the weights can be learned end-to-end.  

\subsection{Injecting temporal memory}

To account for short term memory, we use an intuitive approach (sketched by the red arrows in Fig.~\ref{fig:archi}): we re-inject the class scores obtained at time step $t-1$ into the model learning the features using time step $t$. This type of recursion is typically used in standard recurrent neural networks~\cite{888}, and we use it here in the following way. Since the problem of cloud detection is binary, the class score is two-dimensional (one score per class) and of the same size as the input ($r \times c$). For a model considering $T$ time steps, we make $T$ exact copies of the hypercolumn network presented in Section~\ref{sec:hc}. Then, we modify the input of {the} network by adding two features to the original image by concatenation: such additional input corresponds to the cloud scores map (of size $r \times c \times 2$) obtained at the previous time step $t-1$. For the first time step ($t = 1$) we do not have such scores map: in this case the scores map is replaced by an array of zeros. These simple temporal connections make each time step aware of the prediction of the previous one and force temporal smoothness through time. We avoid forgetting of previous states by using a composite loss function, detailed below.

\vspace{.3cm}

\noindent\textbf{Losses.} Since the model is now a concatenation of $T$ identical CNNs, each prediction task has its own loss function, each one corresponding to a standard cross entropy loss for semantic segmentation:
\begin{equation}
  \mathcal{L}(y,\hat{y}) = - \frac{1}{N} \sum_c \omega_c \llbracket y_i = c \rrbracket \
\log \big(p(y_i = c | x_i)\big).
\label{eq:crossentropy}
\end{equation}
The combination of the losses is a weighted linear combination:
\begin{equation}
\mathcal{L}(y,\hat{y}) = \sum_{t=1}^T \beta^t \mathcal{L}^t(y^t,\hat{y}^t),
\end{equation}
where labels $y^t$ correspond to the cloud map at step $t$, and the $\beta^t$ terms weight the losses by importance. In this case, the importance is increasing the closer we get to the time instance being predicted, since \textit{1)} we want to be accurate in the final prediction (where large temporal information is accumulated) and \textit{2)} we do not want the first loss (at $t = 1$) to become too important (since it uses an empty array of cloud scores as an input). 

\vspace{.3cm}

\noindent\textbf{Parameters update.}
The $T$ CNN models are initialized as identical models and correspond to a single model deployed over time. Since we want {it} to predict correct cloud maps rather than specializing in predicting the mask at a specific time step, the models must remain identical during training. This has several advantages: \textit{1)} the first timestep learns informative filters even though it is not receiving any input from the unavailable cloud scores; \textit{2)} all the time steps classifiers improve after each backpropagation step; \textit{3)} eased backpropagation through the sequence, since a single backpropagation through the whole chain of models will make reaching the first models hard, because of vanishing gradient problems for the last loss~\cite{Hoc01}. 
To do so, we simply average parameter values corresponding to the same filters of the hypercolumn, for all the convolutional (filter weights and bias) and batch normalization filters, at the end of each backpropagation pass. We do not use LSTM units.

\section{Experiments}

\subsection{Data and setup}
We test our network on a dataset provided by EUMETSAT, containing Seviri/MSG Level 1.5 acquisitions acquired in 2010. We considered one landmark location in Dakhla (Western Sahara), which involves
$35'040$ MSG acquisitions (one every 15 minutes over the year) with a fixed resolution of $20 \times 26$ pixels. To be able to train a single model during day and night, we discarded the visible RGB bands and consider only the eight infrared bands of Seviri. 

Level 2 cloud masks were provided~\cite{Der05} and used as the best available `ground truth'. Since {the} ground truth was sometimes incorrect around coastal areas or thermal exchange regions, we masked pixels in a buffer area around them during training, so that the network does not learn spurious patterns related to these inconsistencies. The entire cloud mask was nevertheless considered at test time.

Our network setup is sketched in Fig.~\ref{fig:archi}. 
We trained our networks for a total of 500 epochs, 
using an initial learning rate of $3.5\cdot10^{-5}$, halved after $350$, $400$, $425$ and $450$ epochs, respectively. 

CNN model{s} use the whole images as training samples: in this sense, a random extraction of pixels (as in~\cite{PEREZSUAY201754}) to validate the models is not possible. To avoid temporal positive biases, we use the following train / test strategy: we keep every third month as a separate set of test samples, i.e. the months of March, June, September and December, and use all the data issued from the other eight months as training samples. Each image, corresponding to a $15$ minutes time span, is used in its entirety either to train or test the models. Temporal sequences of length $t = [2,\ldots, 5]$ are considered in the RNN model. 

We compare our proposed RNN with a hypercolumn CNN using only the image at time $t$ to perform the cloud detection (therefore with $t=1$). 
{We train several instances of the RNN with different timesteps, ranging from $t=2$ to $t=5$.} 

\subsection{Results and discussion}
\begin{figure}[!t]
\includegraphics[width=\columnwidth]{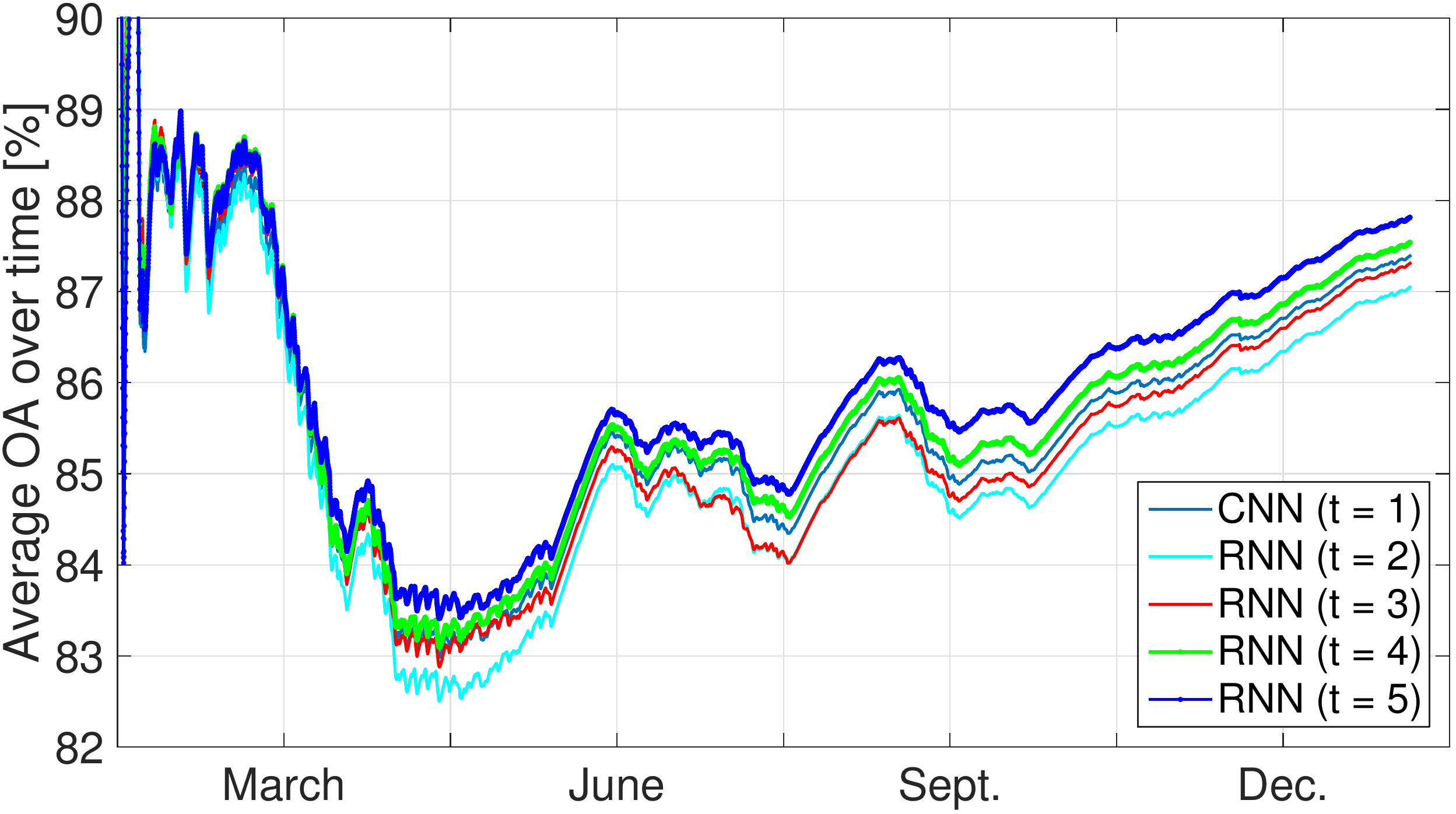}
\caption{Performance of the RNNs in the four test months for different temporal lags $t$.}\label{fig:res}
\end{figure}

Figure~\ref{fig:res} illustrates the numerical comparisons. We used  weighted averaged overall accuracy $$\bar{OA}(i) = \frac{1}{i}\sum_{j=1}^{i} OA(j),$$ 
where $OA(j)$ is the overall accuracy obtained at time instant $j$ preceding the current one. We observe promising performances over all the different months left out for testing. Averaged accuracies span between 82\% and 88\% consistently over the image time series sequence.

The injection of temporal memory improves the model only when sufficient time steps are involved in learning. For instance, for $t=2$ the result slightly worsens the original hypercolumn predictions. We suppose that this behavior is due to the fact that, in the case with $t=1$ the network simply learns to ignore the extra inputs (the $(r \times c \times 2)$ matrix of zeros), while in $t=2$ the weights are updated in an informative way after the first recursion, but the role of the empty inputs is still too strong. With $t=4$ and $t=5$, the RNN makes full use of the predictions obtained at the previous step{s} and provides better predictions, with a moderate +1\% improvement of the prediction over the whole sequence.

\begin{figure*}[!t]
\includegraphics[width = .9\linewidth]{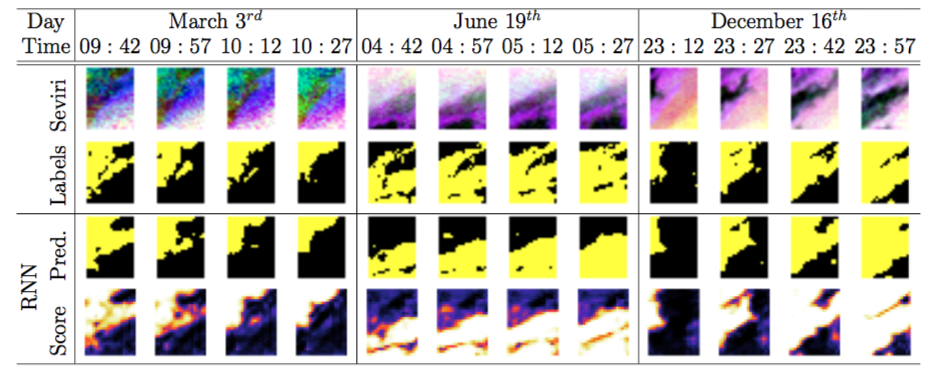}
\caption{Examples of the RNN predictions for the model with $t=5$ in three time instants of the test set.}\label{fig:movie}
\end{figure*}

Figure~\ref{fig:movie} illustrates three parts of the sequences predicted by the RNN with $t=5$: the first, a morning sequence in March, shows that the RNN can follow the global structure of the cloud, but lacks slightly in geometrical accuracy, as it can be seen in the clouds scores obtained by the final classifier. In the second, a night sequence in July, the model predicts correctly the cloud over land, but fails predicting precisely the cloud over the sea. For the last, a night sequence in December, the prediction is very precise and can precisely outline the cloud. Note that both predictions are output{ted} by the exact same model, contrarily to {the divide-and-conquer} strategy  in~\cite{PEREZSUAY201754}.

\section{Conclusions}

In this work, we studied the suitability of a 
convolutional neural network with temporal memory to the prediction of pixel-wise cloud delineation from infrared imagers on meteorological satellites. The 15-minutes sampling rate allows to encode strong temporal information, {which} we injected by using a recurrent neural network structure, where the output of the network at time $t$ becomes an input for the model predicting $t+1$ {and further timesteps}.

We built a single model able to predict clouds accurately day and night, and tested it in a region in the Western Sahara over a 1 year sequence acquired in 2010. The results show that the proposed RNN can predict clouds with an accuracy close to 90\% over the whole sequence, and has no persistent bias related to day/night or the validation month.

\vspace*{-.5mm}

\small{
\bibliographystyle{IEEEtran}
\bibliography{bib}
}


\end{document}